\documentstyle[12pt]{article}
\newtheorem{thm}{Theorem}[subsection]
\newtheorem{cor}[thm]{Corollary}
\newtheorem{lem}[thm]{Lemma}

\newtheorem{defn}[thm]{Definition}

\begin{document}
\input{amssym}
\renewcommand{\theequation}{\arabic{section}.\arabic{equation}}
\def\di{\displaystyle}
\title{Lie Symmetries and Solutions of KdV
Equation}
\author{Mehdi Nadjafikhah\thanks{Department of Mathematics, Iran University of Science and Technology,
Narmak, Tehran, I.R.Iran. e-mail: m\_ nadjafikhah@iust.ac.ir} \and
Seyed-Reza Hejazi\thanks{e-mail: reza\_ hejazi@iust.ac.ir}}
\date{February 1, 2008}
\maketitle
\begin{abstract}
Symmetries of a differential equations is one of the most
important concepts in theory of differential equations and
physics. One of the most prominent equations is KdV (Kortwege-de
Vries) equation with application in shallow water theory. In this
paper we are going to explain a particular method for finding
symmetries of KdV equation, which is called Harrison method. Our
tools in this method are Lie derivatives and differential forms,
which will be discussed in the first section more precisely. In
second chapter we will have some analysis on the solutions of KdV
equation and we give a method, which is called first integral
method for finding the solutions of KdV equation.
\end{abstract}


{\rm A.M.S. 2000 Subject Classification:} {\em 35Bxx, 58Axx,
27E70, 34A26, 22E60.}
\section*{Introduction}
There are diverse methods for finding symmetries of differential
equations. One of the most important one is Lie method. In this
method we need a lot of basis, such as Lie group theory,
prolongation and ..., which are foundations of Lie's theory of
symmetry groups of differential equations. This is a firm method
for finding symmetries and it has a lot of applications in
differential equations theory. Harrison's method is an another
one, which does not need the above necessaries. Here we act this
method on the KdV equation, a very famous differential equation,
and testing the results, specially with the Lie method see
\cite{[4]},\cite{[14]}, \cite{[15]}, \cite{[16]}.
\section{Harrison's Method}
The method proceeds as follows. We consider a set of partial
differential equations, defined on a differentiable manifold $E$
with $p$ independent variables and $q$ dependent variables, here
we assume $p=2$ and $q=1$. We define the partial derivatives of
the dependent variables as new variables (prolongation) in
sufficient number to write the equation as second order equation,
then we can construct a set of differential forms. We speak of the
set of forms, representing the equations, as an ideal $I$. It is
to be closed. (See \cite{[6]}, \cite{[7]}, \cite{[11]} for more
details.)
\par
In this method Lie derivative of a form with respect to a vector
field makes an important roll. Lie derivatives of geometrical
objects, like differential forms, are associated with symmetries
of those objects. The Lie derivative of a geometrical object
carries it along a path, determined by a vector field \textbf{v},
in its manifold. If the Lie derivative vanishes, then the vector
\textbf{v} represents the direction of an infinitesimal symmetry
in the manifold. We may construct the Lie derivative
$(\mathcal{L})$ of differential form in the ideal $I$.
\par
Setting the Lie derivative of these forms equal to zero should
therefore represent symmetries. We require that the Lie
derivatives of the forms in $I$ to be linear combinations of those
forms themselves and when they vanish they Lie derivatives also
vanish. We can illustrate this by writing ${\cal
L}_{\textbf{v}}I\equiv 0\;\;\;\;\mbox{mod I},$ or ${\cal
L}_{\textbf{v}}I\subset I,$ for Lie derivative properties see
\cite{[12]}.
\subsection{Construction of Differential Forms}
First of all substitute $u_x$ by $w$, thus the KdV equation
$u_{xxx}+uu_{x}+u_{t}=0$, reduced to the following second order
partial differential equation
\begin{eqnarray}
w_{xx}+uw+u_{t}=0. \label{eq:1.1}
\end{eqnarray}
The jet space corresponds to the equation (\ref{eq:1.1}) is a
$8-$dimensional manifold with the coordinate
$(t,x,u,u_{t},u_{x},u_{tt},u_{tx},u_{xx}).$ Corresponding to the
equation we have the following contact 1-forms
\begin{eqnarray}
\theta^{1} &=& du-u_{t}dt-u_{x}dx,\nonumber \\
\theta^{2} &=& du_{t}-u_{tt}dt-u_{tx}dx,\nonumber\\
\theta^{3} &=& du_{x}-u_{tx}dt-u_{xx}dx.\nonumber
\end{eqnarray}
Consequently our required forms are
\begin{eqnarray*}
\alpha^{1} &=& \theta^{1}\wedge \theta^{2} =
(u_{x}u_{tt}-u_{t}u_{tx})dx\wedge dt+u_{tx}dx\wedge
du-u_{x}dx\wedge du_t\\
&&+u_{tt}dt\wedge du-u_{t}dt\wedge du_{t}+du\wedge du_{t},\\
\alpha^{2} &=& \theta^{1}\wedge \theta^{3} =
(u_{x}u_{tx}-u_{t}u_{xx})dx\wedge dt+u_{xx}dx\wedge
du-u_{x}dx\wedge du_{x}\\
&& +u_{tx}dt\wedge du-u_{t}dt\wedge du_{x}+du\wedge du_{x},\\
\alpha^{3} &=& \theta^{2}\wedge \theta^{3} =
(u_{tx}^{2}-u_{tt}u_{xx})dx\wedge dt+u_{xx}dx\wedge
du_{t}-u_{tx}dx\wedge du_{x}\\
&& +u_{tx}dt\wedge du_{t}-u_{tt}dt\wedge du_{x}+du_{t}\wedge
              du_{x},\\
\alpha^{4} &=& d\theta^{1} = dx\wedge du_{x}+dt\wedge du_{t},\\
\alpha^{5} &=& d\theta^{2} = dx\wedge du_{tx}+dt\wedge du_{tt},\\
\alpha^{6} &=& d\theta^{3} = dx\wedge du_{xx}+dt\wedge du_{tx},\\
\alpha^{7} &=& udt\wedge du-dt\wedge du_{xx}-dx\wedge du.
\end{eqnarray*}
Consider a vector field on the assumed jet space in the form of
\begin{eqnarray*}
\textbf{v} = A_{1}\frac{\partial}{\partial t}
+A_{2}\frac{\partial}{\partial x} +A_{3}\frac{\partial}{\partial
u} +A_{4}\frac{\partial}{\partial u_{t}}+
            A_{5}\frac{\partial}{\partial u_{x}}+A_{6}\frac{\partial}{\partial u_{tt}}+A_{7}\frac{\partial}{\partial u_{tx}}+A_{8}\frac{\partial}{\partial
           u_{xx}},
\end{eqnarray*}
where $A_{i}$s, $i=1,...,8$, are smooth functions of
$(t,x,u,u_{t},u_{x},u_{tt},u_{tx},u_{xx})$.
\par
The next step is solving the following huge!! partial differential
equations system,
\begin{eqnarray}
{\cal L}_{\textbf{v}}\alpha^{i}=\lambda_{i}\alpha^{i},\;\;\;\;
\mbox{for}\; i=1,2,3,\label{eq:1.2}
\end{eqnarray}
 for smooth functions $\lambda_{i}$.

The system (\ref{eq:1.2}) has the form
\begin{eqnarray*}
& \di -\Big(\frac{\partial}{\partial
u_{tt}}A_{3}\Big)+\cdots+\Big(\frac{\partial}{\partial
u_{xx}}A_{1}\Big)u_{tx}^{2}=0, & \\
& \di -u\Big(\frac{\partial}{\partial
u_{xx}}A_{2}\Big)u_{x}u_{tx}+\cdots-\Big(\frac{\partial}{\partial u_{x}}A_{5}\Big)u_{tx}=0, & \\
& \vdots& \\
& \di -\Big(\frac{\partial}{\partial
t}A_{2}\Big)+\cdots-\Big(\frac{\partial}{\partial
u_{xx}}A_{7}\Big)=0, & \\
& \di -\Big(\frac{\partial}{\partial
u_{xx}}A_{4}\Big)+\cdots+\Big(\frac{\partial}{\partial
u_{tx}}A_{2}\Big)u_{tx}=0. &
\end{eqnarray*}
After solving this system with respect to $A_{1},...,A_{8}$, we
have
\begin{eqnarray*}
\begin{array}{lll}
\di A_{1}=c_{1}t+c_{2}, & \di A_{2}=c_{1}x/3+c_{3}t+c_{4},& \di
A_{3}=-2c_{1}u/3+c_{3}, \\
\di A_{4}=-c_{3}u_{x}-5c_{1}u_{t}/3,& \di A_{5}=-c_{1}u_{x}, &
\di A_{6}=-2c_{3}u_{tx}-8c_{1}u_{tt}/3,\\
\di A_{7}=-c_{3}u_{xx}-2c_{1}u_{tx}, & A_{8}=-4c_{1}u_{xx}/3,
\end{array}
\end{eqnarray*}
where $c_{1}, c_{2}, c_{3}, c_{4}$ are arbitrary constants, thus
\begin{eqnarray*}
\textbf{v}=(c_{1}t+c_{2})\frac{\partial}{\partial
             t}+(\frac{c_{1}}{3}x+c_{3}t+c_{4})\frac{\partial}{\partial
             x}+(c_{3}-\frac{2c_{1}}{3}u)\frac{\partial}{\partial
             u}-(c_{3}u_{x}+\frac{5c_{1}}{3}u_{t})\frac{\partial}{\partial
             u_{t}}\\
          -c_{1}u_{x}\frac{\partial}{\partial
             u_{x}}-(2c_{3}u_{tx}+\frac{8c_{1}}{3}u_{tt})\frac{\partial}{\partial
             u_{tt}}-(c_{3}u_{xx}+2c_{1}u_{tx})\frac{\partial}{\partial
             u_{tx}}-\frac{4c_{1}}{3}u_{xx}\frac{\partial}{\partial
             u_{xx}}.
\end{eqnarray*}
We claim that the four vector fields which will be constructed
from \textbf{v}, make four dimensional symmetry group for KdV
equation such as we can obtain in Lie method.
\begin{eqnarray*}
\textbf{v}_{1}&=&\frac{\partial}{\partial x},\;\;\;\;\;\textbf{v}_{2}=\frac{\partial}{\partial t},\\
\textbf{v}_{3}&=&t\frac{\partial}{\partial
                 x}+\frac{\partial}{\partial u}-u_{x}\frac{\partial}{\partial
                 u_{t}}-2u_{tx}\frac{\partial}{\partial
                 u_{tt}}-u_{xx}\frac{\partial}{\partial u_{tx}},\\
\textbf{v}_{4}&=&x\frac{\partial}{\partial
                 x}+3t\frac{\partial}{\partial t}-2u\frac{\partial}{\partial
                 u}-5u_{t}\frac{\partial}{\partial
                 u_{t}}-3u_{x}\frac{\partial}{\partial
                 u_{x}}\\
              & &-8u_{tt}\frac{\partial}{\partial
                u_{tt}}-6u_{tx}\frac{\partial}{\partial
                u_{tx}}-4u_{xx}\frac{\partial}{\partial u_{xx}}.
\end{eqnarray*}
First of all we should show that the set of these four vector
fields
$\{\textbf{v}_{1},\textbf{v}_{3},\textbf{v}_{3},\textbf{v}_{4}\}$,
makes a Lie algebra construction, it is sufficient to show that
$[\textbf{v}_{i},\textbf{v}_{j}]$ lies in the vector space
constructed by
$\{\textbf{v}_{1},\textbf{v}_{2},\textbf{v}_{3},\textbf{v}_{4}\}$,
where [ , ] is the Lie bracket of vector fields. By a directly
computation we have
\begin{eqnarray*}
\begin{array}{lll}
{[\textbf{v}_{1},\textbf{v}_{2}]}=0,             &[\textbf{v}_{1},\textbf{v}_{3}]=0,
&[\textbf{v}_{1},\textbf{v}_{4}]=\textbf{v}_{1},\\[2mm]
{[\textbf{v}_{2},\textbf{v}_{3}]}=\textbf{v}_{1},&[\textbf{v}_{2},\textbf{v}_{4}]=3\textbf{v}_{2},&[\textbf{v}_{3},\textbf{v}_{4}]=-2\textbf{v}_{3},
\end{array}
\end{eqnarray*}
if we label the point parts of these vector fields by
\begin{eqnarray*}
\begin{array}{lll}
\di X_{1}=\frac{\partial}{\partial x}, &&
\di X_{2}=\frac{\partial}{\partial t},\\[3mm]
\di X_{3}=t\frac{\partial}{\partial x}+\frac{\partial}{\partial
u}, && \di X_{4}=x\frac{\partial}{\partial
x}+3t\frac{\partial}{\partial t}-2u\frac{\partial}{\partial u},
\end{array}
\end{eqnarray*}
it is easy to see that the third prolongations of
$X_{1},X_{2},X_{3},X_{4}$, vanishes the KdV equation, (see
\cite{[14]}, \cite{[15]}, for more details.) $\di
X_{i}^{(3)}(u_{xxx}+uu_{x}+u_{t})=0,$ for $i=1,2,3,4$. Where
$X_{i}^{(3)}$, is the third prolongation of $X_{i}$, thus
$\{X_{1},X_{2},X_{3},X_{4}\}$, makes a set of four parameter
symmetry group for KdV equation. It is necessary to say that the
four vector fields
$\textbf{v}_{1},\textbf{v}_{2},\textbf{v}_{3},\textbf{v}_{4}$,
which we had found are second prolongation of
$X_{1},X_{2},X_{3},X_{4}$, because in the incipience of the
section, we decreased the order of equation to two.
\section{Solutions and First Integral of KdV Equation}
First integral(s) of an ordinary differential equation, which will
be defined more precisely, is a function which its differential
lies in the annihilator of the distribution corresponds to the
assumed ODE. If we have a first integral(s) of an ODE we can find
its solution, specially for some differential equations which we
are unable to find its solution with any model. For the beginning
of this part, we have some analysis on solutions of equations
then, we should have a view on some fundamental concepts in theory
of differential equations. The reader may see \cite{[13]} for more
details. Now consider the KdV equation, let $G$ be an
$n-$dimensional Lie group with the Lie algebra $\goth g$. $G$ is
called \textit{solvable} if there exist a sequence of subgroups
$\{e\}\subset G_{0}\subset G_{1}\subset \cdots \subset
G_{n-1}\subset G_{n}=G,$ such that each $G_{i}$ is a normal
subgroup of $G_{i+1}$. This is equivalence to the requirement that
the corresponding subalgebras of $\goth g$ satisfies $[{\goth
g}_{i},{\goth g}_{i+1}]\subset {\goth g}_{i}.$
\subsection{Bianchi Theorem}
A theorem of Bianchi, \cite{[3]}, states that if an ordinary
differential equation admits an $n-$dimensional solvable symmetry
group, then its solution can be determined, by quadratures, from
those to reduces the order of equation; see also \cite{[15]}.
\begin{thm}\label{theorem:2.3.1}
Let $\Delta(x,u^{(m)})=0$ be an mth order ordinary differential
equation. If $\Delta$ admits a solvable n-parameter group of
symmetries G such that for $1\leq i\leq n$ the orbits of $G^{(i)}$
(ith prolongation of G) have dimension i, then the general
solution of $\Delta$ can be found by quadratures from the general
solution of an (m-n)th order differential equation
$\tilde{\Delta}(y,w^{(m-n)})=0.$ In particular, if $\Delta$ admits
an m-parameter solvable group of symmetries, then the general
solution to $\Delta$ can be found by quadratures alone.
\end{thm}
The Bianchi theorem is extended on an arbitrary distribution,
which theorem (\ref{theorem:2.3.1}) is its special case, \cite{[13]}.\\
By a change of variables as $y = x-ct$ and $v = u$, the reduced
equation is $v'''+vv'-cv'=0,$ where $v'''=d^{3}v/dy^{3}$ and
$v'=dv/dy$. This can be immediately integrated once,
$v''+v^{2}/2-cv=c_{1}$, another integration reduced equation to
\begin{eqnarray}
\frac{1}{2}{v'}^{2}+\frac{1}{6}v^{3}-\frac{1}{2}cv^{2}-c_{1}v-c_{2}=0,\label{eq:1.7}
\end{eqnarray}
where $c_{1}$ and $c_{2}$ are constants. The general solution can
be written in terms of elliptic function, $u={\cal
P}(x-ct+\varepsilon)$, $\varepsilon$ being an arbitrary phase
shift. If $u\rightarrow 0$ sufficiently rapidly as
$|x|\rightarrow\infty$, then $c_{1}=c_{2}=0$ in equation
(\ref{eq:1.7}), this equation has the solution
$v=3c\;\textrm{sech}^{2}\Big[\sqrt{c}/2+\varepsilon\Big],$
provided the wave speed $c$ is positive. This produce the
celebrated "one soliton" solutions
$$u(x,t)=3c\mbox{sech}^{2}\Big[\frac{1}{2}\sqrt{c}(x-ct)+\varepsilon\Big],$$
to the KdV equation, which is called \textit{Travelling Wave
Solution.}
\par
There is a lot kinds of solutions for Kortweg-de Vries, such as
\textit{Galilean-Invariant Solution} and \textit{Scale-Invariant
Solution}.
\par
Consider the flows generated by the four symmetries,
\begin{eqnarray*}
\begin{array}{lcl}
\di \theta_{1}(s)(t,x,u)=(t,x+s,u), && \di
\theta_{2}(s)(t,x,u)=(t+s,x,u),\\[2mm]
\di \theta_{3}(s)(t,x,u)=(t,x+ts,s+u),&&
\di \theta_{4}(s)(t,x,u)=(te^{3s},xe^{s},ue^{-2s}),\\
\end{array}
\end{eqnarray*}
according to the flows, we can find some special solutions for KdV
equation. For example if $H(x,t)$ be a solution of KdV, then is so
$u(t,x)=\delta^{2}H(\delta^{3}t+\alpha,\delta x+\beta +\gamma
\delta t)-\lambda,$ where $\alpha$, $\beta$, $\gamma$, $\delta$
and $\lambda$ are arbitrary constants. We can see that
\begin{eqnarray*}
u(t,x)&=&\frac{12\gamma^{2}}{(\beta t+\gamma x+\alpha)^2}-\frac{\beta}{\gamma},\\
u(t,x)&=&-12\mbox{tanh}(x^2)+8,\\
u(t,x)&=&-12\gamma^{2}\tanh(\beta t+\gamma
x+\alpha)^2+8\gamma^{2}-\frac{\beta}{\gamma},
\end{eqnarray*}
are another solutions which are constructed with the symmetries
and flows.
\par
If $u$ depends only on $x$, i,e., $u=y(x)$, then KdV equation
reduced to
\begin{eqnarray}
y'''+yy'=0,\label{eq:1.8}
\end{eqnarray}
and $X_{1}=\partial/\partial x$, $X_{2}=x\partial/\partial
x-2y\partial/\partial y$, are two symmetries for equation
(\ref{eq:1.8}), and $x=v(u)$, $y=u$, is a change of variable
corresponds to $X_{1}$ for equation (\ref{eq:1.8}). With these new
variables and by substituting $v'=\eta(u)$ the equation
(\ref{eq:1.8}) changes to
\begin{eqnarray}
u\eta^{4}+3{\eta'}^{2}-\eta\eta''=0,\label{eq:1.10}
\end{eqnarray}
where $\eta'=d\eta/du$. This new equation (\ref{eq:1.10}) has the
following three parameter symmetry group
\begin{eqnarray*}
\tilde{X_{1}}=\eta^{3}\frac{\partial}{\partial \eta},\;\;\;
\tilde{X_{2}}=\eta^{3}u\frac{\partial}{\partial \eta},\;\;\;
\tilde{X_{3}}=2u\frac{\partial}{\partial
u}-3\eta\frac{\partial}{\partial \eta},
\end{eqnarray*}
corresponding to $\tilde{X_{3}}$, we have the change of variable
\begin{eqnarray*}
t =\eta(u)u^{\frac{3}{2}},\;\;\;\; s(t)=\frac{1}{2}\ln(u),
\end{eqnarray*}
by substituting $s'=\xi(t)$, equation (\ref{eq:1.10}) reduces to
\begin{eqnarray}
4t^{4}\xi^{3}+3\xi-10t\xi^{2}+12t^{2}\xi^{3}+t\xi'=0,\label{eq:1.11}
\end{eqnarray}
where $\xi'=d\xi/dt$. This equation has one parameter symmetry
group
$$\bar{X}=t(3+t^2)\frac{\partial}{\partial t}-3(1+t^{2})\xi\frac{\partial}{\partial \xi}.$$
The solution of the equation (\ref{eq:1.11}) is the general
solution of KdV equation.
\subsection{Frobenius Theorem}
The Frobenius theorem is one of the most important theorem in
theory of differential equations, which one of its results
guarantees that an ODE has first integral(s).
\begin{defn}
Suppose $M$ is an $n-$dimensional manifold and $p\in M$. A choice
of $k-$dimensional linear subspace $D_{p}\subset T_{p}M$ is called
a $k-$dimensional tangent distribution or a $k-$dimensional
distribution. $D_{p}$ is called smooth if $D=\bigcup_{p\in
M}D_{p}\subset TM,$ is a smooth subbundle of $TM$.
\end{defn}
Here $T_{p}M$ and $TM$ are tangent space on $M$ in point $p$, and
$TM$ is the tangent bundle on $M$.
\begin{lem}\cite{[12]}\label{lem:2.2.1}
Let $M$ be a smooth $n-$manifold, and let $D\subset TM$  be a
$k-$dimensional distribution. Then $D$ is smooth if and only if
each point $p\in M$ has a neighborhood $U$ on which there are
smooth 1-forms $\omega^{1},...,\omega^{n-k}$ such that for each
$q\in U$,
$$D_{q}=\ker\omega^{1}\Big|_{q}\cap\cdots \cap \ker\omega^{n-k}\Big|_{q}.$$
\end{lem}
More precisely, if we denote the annihilator of $D$ by
$$\textrm{Ann}(D)=\Big\{\omega\in \Omega^{1}(M): \omega=0\; \mbox{on}\; D\Big\},$$
then for any $\omega^{i}$ defined in lemma (\ref{lem:2.2.1}) we
have $\omega^{i}\in \textrm{Ann}(D).$
\par
If $D$ is a distribution generates by
$\{\textbf{v}_{1},\cdots,\textbf{v}_{k}\}$, then $D$ could be
discussed by $\{\omega^{1},...,\omega^{n-k}\}$ too. We will show
such a distribution as
$$D={\cal F}(\textbf{v}_{1},...,\textbf{v}_{k})={\cal F}(\omega^{1},...,\omega^{n-k}),$$
where $n-k$ is codimension of $D$.
\begin{defn}
Suppose $D={\cal F}(\textbf{v}_{1},...,\textbf{v}_{k})$ is a
$k-$dimensional distribution. The distribution $D^{(1)}$ which is
generates by the vector fields
$\{\textbf{v}_{1},...,\textbf{v}_{k}\}$ and by all possible sorts
of commutators $[\textbf{v}_{i},\textbf{v}_{j}]
(i<j;i,j=1,...,k)$, is called the \textit{first derivative} of
$D$, i.e.,
$$D^{(1)}={\cal F}(\textbf{v}_{1},...,\textbf{v}_{k},[\textbf{v}_{1},\textbf{v}_{2}],...,[\textbf{v}_{1},\textbf{v}_{k}],...,
[\textbf{v}_{k-1},\textbf{v}_{k}]).$$
\end{defn}
\begin{lem}\cite{[13]}
Let $D={\cal F}(\textbf{v}_{1},...,\textbf{v}_{k})$ is a
distribution such that ${\cal F}(\omega^{1},...,\omega^{n-k})$.
Then $D^{(1)}=D$ if and only if for $i=1,...,n-k$
\begin{eqnarray}
d\omega^{i}\wedge \omega^{1}\wedge ...\wedge
\omega^{n-k}=0.\nonumber
\end{eqnarray}
\end{lem}
\begin{defn}
A smooth distribution $D$ on a smooth manifold $M$ is called
\textit{Completely Integrable Distribution} or a CID, if all
points of $M$ contain in an integral manifold of $D$. (A
submanifold $N\subset M$ is called an integral manifold of $D$ if
$T_{p}N\subset D$.)
\end{defn}
Now we are ready to give the Frobenius theorem.
\begin{thm}\cite{[12]}
Let $D$ is a smooth distribution on a smooth manifold $M$. $D$ is
CID if and only if $D^{(1)}=D$.
\end{thm}
\par
Now by using the Frobenius theorem it will be shown that any first
order ODE has first integral(s). This is the result of  the
Frobenius theorem and its following corollary, before we have a
necessary definition.
\begin{defn}
Let $D$ is a smooth distribution on smooth manifold $M$, a  smooth
function $\varphi\in C^{\infty}(M)$ is called a first integral for
$D$ if $d\varphi\in \textrm{Ann}(D)$. Here $C^{\infty}(M)$ is the
set of all smooth real valued function on manifold $M$. Another
definition for a CID, is that $D$ is CID if and only if there
exist $n-k$ first functional independent integrals
$\varphi_{1},...,\varphi_{n-k}$ such that $D={\cal
F}(d\varphi_{1},...,d\varphi_{n-k}),$
\end{defn}
\noindent so if a distribution is CID it means that it has first
integral(s).
\begin{cor}
Suppose $D$ is a smooth distribution such that
$D=\mathcal{F}(\omega)$. Then $D$ is CID if and only if
\begin{eqnarray}
\omega\wedge d\omega =0.\label{eq:1.3}
\end{eqnarray}
Now consider a first order ODE
\begin{eqnarray}
y'=\frac{dy}{dx}=f(x,y),\label{eq:1.4}
\end{eqnarray}
\end{cor}
it's clear that the equation (\ref{eq:1.4}) obtained by taking the
1-form $\omega=dy-f(x,y)dx,$ to zero, so $\omega$ satisfies the
equation (\ref{eq:1.3}), thus the distribution corresponds to the
equation (\ref{eq:1.4}) has first integral(s).
\subsection{Symmetries of Distribution}
\begin{defn}
A symmetry of a distribution is the transformation of the manifold
$M$ that maps distribution into itself. In other words, a
diffeomorphism $F:M\rightarrow M$ is a \textit{symmetry of a
distribution D} if $F_{*}(D_{p})=D_{F(p)},$ for all $p\in M$. Here
$F_{*}$, is push forward of $F$.
\end{defn}

Suppose \textbf{v} is an smooth vector field on manifold $M$, and
$\theta_{t}$ be its flow, then we know $\theta_{t}:M\rightarrow M$
induces a diffeomorphism on $M$.
\begin{defn}
$\theta_{t}$ which has defined above is called an
\textit{infinitesimal symmetry} or a \textit{symmetry} of a
distribution $D$ if $\theta_{t}$ along the vector field \textbf{v}
consists of symmetries of $D$. i.e,
$\theta_{*}(D_{p})=D_{\theta_{t}(p)},$ for all $p\in M$ and $t$.
\end{defn}
Denote $\textrm{Sym}(D)$  the set of all infinitesimal symmetries
of distribution $D$.
\begin{thm}\cite{[13]}
Let $D$ be a distribution and $\mathcal{X}(D)$ denotes the set of
all vector fields on $D$, then the following statements are
equivalent.
\begin{itemize}
\item[i)] \textbf{v} $\in \textrm{Sym}(D)$.
\item[ii)] $\forall \textbf{w}\in \mathcal{X}(D)\Rightarrow
[\textbf{v},\textbf{w}]\in \mathcal{X}(D).$
\item[iii)] $\forall \omega\in \textrm{Ann}(D)\Rightarrow
{\cal L}_{\textbf{v}}\omega \in \textrm{Ann}(D).$
\end{itemize}
\end{thm}
In this theorem $\mathcal{L}_{\textbf{v}}\omega$, denotes the Lie
derivative of $\omega$ with respect to \textbf{v}.
\begin{cor}
$\textrm{Sym}(D)$ has a real Lie algebra structure with respect to
the commutator of the vector fields.

\end{cor}
\subsection{Distribution with a Commutative Symmetry Algebra}
Let $\goth g$  be a commutative symmetry Lie algebra which its
dimension is equal to the dimension of ${\cal
F}(\omega^{1},...,\omega^{n-k})$ is equal to $k$. Let
$\{\textbf{v}_{1},...,\textbf{v}_{k}\}$ be a basis of $\goth g$
and let $D$ is a CID. Then form the matrix
\begin{eqnarray*}
\textbf{Z}=\left(\begin{array}{ccc}\omega^{1}(\textbf{v}_{1})&\cdots
&\omega^{1}(\textbf{v}_{k})\\
\vdots &\vdots &\vdots\\
\omega^{k}(\textbf{v}_{1})&\cdots
&\omega^{k}(\textbf{v}_{k})\end{array}\right),
\end{eqnarray*}
because of independence of $\omega^{i}$s, then $\textbf{Z}^{-1}$
is exist, now we are going to construct a new basis
$\bar{\omega}^{1},...,\bar{\omega}^{k}$ for $\textrm{Ann}(D)$.
These basis are constructed by the following relation
\begin{eqnarray}
\left(\begin{array}c\bar{\omega}^{1}\\
\vdots\\
\bar{\omega}^{k}\end{array}\right)=\textbf{Z}^{-1}\left(\begin{array}c\omega^{1}\\
\vdots\\
\omega^{k}\end{array}\right).\label{eq:1.6}
\end{eqnarray}
It is possible to see that $\bar{\omega}^i$s are closed, so the
functions $\varphi_{i}(p)=\int_{\alpha} \bar{\omega}^i,$ are
called \textit{first integrals} of $D$. Here $\alpha$ is a path
from the fix point $p_{0}$ to a point $p\in M$. Because of
closeness of $\bar{\omega}^{i}$s, the integration is independent
from choice of $\alpha$.
\par
Consider a completely integrable distribution
$D=\mathcal{F}=(\omega)$ with a symmetry \textbf{v}, then
according to the relation (\ref{eq:1.6}), the 1-form
$\bar{\omega}=\frac{1}{\omega(\textbf{v})}\omega,$ is closed and
the function
\begin{eqnarray}
\varphi=\int_{\alpha}\frac{\omega}{\omega(\textbf{v})},\label{eq:1.12}
\end{eqnarray}
is a first integral of $D$.
\par
For another example, consider an ODE in the form of equation
(\ref{eq:1.4}), we know that the corresponding 1-form is
$\omega=dy-f(x,y)dx,$ suppose that
$\textbf{v}=a(x,y).\partial/\partial x+b(x,y).\partial/\partial
y,$ is a symmetry of the above distribution, that is the symmetry
of the equation (\ref{eq:1.4}), $\textbf{Z}=b(x,y)-f(x,y)a(x,y)$
and the differential 1-form
$\bar{\omega}=(dy-f(x,y)dx)/(b(x,y)-f(x,y)a(x,y)),$ is closed, and
the function
$$\varphi=\int_{\alpha}\bar{\omega},\label{eq:1.9}$$
is a first integral of the equation (\ref{eq:1.4}). The function
$\textbf{Z}^{-1}$ is called an \textit{integrating factor} for the
equation (\ref{eq:1.4}).
\subsection{First Integral of KdV Equation}
After we reduced the equation to an ODE, we can obtain its first
integral due to equation (\ref{eq:1.12}). The equation
(\ref{eq:1.7}) could be written as 1-form
$$\omega=dv-\Big(\sqrt{cv^{2}+2c_{1}v+2c_{2}-v^{3}/3}\Big)dy,$$
such that its independent variable $y$, does not enter explicitly,
thus it is obvious that $\textbf{v}=\partial/\partial y$, is a
symmetry for the equation (\ref{eq:1.7}), consequently the closed
$1-$form
$\bar{\omega}=-\omega/\sqrt{+cv^{2}+2c_{1}v+2c_{2}-v^{3}/3},$ is
constructed and the function $\varphi=\int_{\alpha}
\Big(dy-dv/\sqrt{{+cv^{2}+2c_{1}v+2c_{2}}-v^{3}/3}\Big),$ which is
a line integral on any arbitrary path $\alpha$ is the first
integral ($d\varphi\in \textrm{Ann}(D)$ where
$D=\mathcal{F}(\omega)=\mathcal{F}(\textbf{v})$) for KdV equation,
and the set $\big\{u(t,x):\varphi=0\big\},$ is the set of all
solutions for the equation.


\begin{thebibliography}{9}
%
\bibitem{[1]}{\sc Anderson, I.M., Kamran, N., Olver, P.J.},
{\em Internal, External and Generalized Symmetries}, Adv. in Math.
100 (1993), 53-100.
\bibitem{[2]}{\sc Arnol'd, V.L.}, {\em Geometrical Methods in
the Theory of Ordinary Differential Equations}, Springer Verlage,
New York, 1983.
\bibitem{[3]}{\sc Bianchi, L.}, {\em Lezioni sulla Teoria dei
Gruppi Continui Finiti di Transformazioni}, Enrico Spoerri, Pisa,
1918.
\bibitem{[4]}{\sc Bluman, G.W., and Kumei, S.}, {\em Symmetries
and Differential equations}, Springer Verlage, New York, 1989.
\bibitem{[5]}{\sc Burke, W.L.}, {\em Applied Differential
Geometry}, Cambridge University Press, Cambridge, 1985.
\bibitem{[6]}{\sc Harrison B. Kent}, {\em The Differential Form
Method for Finding Symmetries}, SIGMA, Vol. 1 (2005), Paper 001,
12 pages
\bibitem{[7]}{\sc Harrison, B. Kent, Estabrook, F.B.},
{\em Geometric Approach to Invariance Groups and Solution of
Partial Differential System}, J. Math. Phys., 1971, V. 12,
653-666.
\bibitem{[8]}{\sc Humphreys, J.E.}, {\em Introduction to Lie
Groups and Lie Algebras}, Graduate Texts in Mathematics, Vol. 9,
Springer Verlage, New York, 1976.
\bibitem{[9]}{\sc Ibragimov, N.H.}, {\em Group Analysis of
Ordinary Differential Equations and the Invariance Principle in
Mathematical Physics}, (for teh 150th anniversary of Sophus Lie),
Russian Math, Surveys 47:4 (1992).
\bibitem{[10]}{\sc Ibragimov, N.H.}, {\em CRC HandBook of Lie Group
Analysis of Differential Equations}, Vol. 1, CRC Press, Boca
Raton, F1., 1994.
\bibitem{[11]}{\sc Kersten, P.H.M.}, {\em Infinitesimal Symmetries:
A Computational Approach}, Ph.D. Thesis, Twente University of
Technology, Enschede, The Netherlands, 1985.
\bibitem{[12]}{\sc Lee, John, M.}, {\em Introduction to Smooth
Manifolds}, Springer Verlage, New York, 2002.
\bibitem{[13]}{\sc Kushner, A,. Lychagin, V. and Robstov, V.},
{\em Contact Geometry and Non-Linear Differential Equations},
Cambridge University Press, Cambridge, 2007.
\bibitem{[14]}{\sc Olver, P.J.}, {\em Equivalence, Invariant and
Symmetry}, Cambridge University Press, Cambridge 1995.
\bibitem{[15]}{\sc Olver, P.J.}, {\em Applications of Lie Groups to
Differential equations}, Second Edition, GTM, Vol. 107, Springer
Verlage, New York, 1993.
\bibitem{[16]}{\sc Ovsiannikov, L.V.}, {\em Group Analysis of
Differential Equations}, Academic Press, New York, 1982.
\bibitem{[17]}{\sc Schutz, B.F.}, {\em Geometrical Methods of
Mathematical Physics}, Cabridge, Cabmridge University Press, 1980.
\end{thebibliography}
\end{document}